\documentclass{aa}  

\usepackage{graphicx}
\usepackage{natbib}
\usepackage{txfonts}
\usepackage[colorlinks=true,citecolor=blue,linkcolor=blue]{hyperref}

\usepackage{changes}

\graphicspath{{./}{figures/}}
\begin{document}

   \title{Mass limits of the extremely fast-spinning white dwarf CTCV J2056--3014}

   \author{Edson Otoniel
          \inst{1}
          \and
          Jaziel G. Coelho\inst{2,}\inst{3}
          \and Sílvia P. Nunes\inst{4}
          \and Manuel Malheiro\inst{4}
          \and Fridolin Weber\inst{5,}\inst{6}
          }

   \institute{Instituto de
  Forma\c{c}\~ao de Educadores, Universidade Federal do Cariri,
  R. Oleg\'ario Emidio de Araujo, s/n - Aldeota, 63260-000 Brejo
  Santo, CE, Brazil\\
\email{edson.otoniel@ufca.edu.br}
         \and
             Departamento de F\'isica,
  Universidade Tecnol\'ogica Federal do Paran\'a, 85884-000
  Medianeira, PR, Brazil\and
  Divis\~ao de Astrof\'isica,
  Instituto Nacional de Pesquisas Espaciais, Avenida dos Astronautas
  1758, 12227--010 S\~ao Jos\'e dos Campos, SP, Brazil\\
\email{jazielcoelho@utfpr.edu.br}
             \and 
Departamento de F\'isica, Instituto Tecnol\'ogico de Aeron\'autica, Pra\c{c}a Marechal Eduardo Gomes, 50 - Vila das Acacias, 12228--900 S\~ao Jos\'e dos Campos, SP, Brazil
\and
  Department of Physics, San Diego
  State University, 5500 Campanile Drive, San Diego, California 92182,
  USA\and
  Center for Astrophysics and Space Sciences,
  University of California at San Diego, La Jolla, California 92093,
  USA\\
 \email{fweber@sdsu.edu }
             }

  \abstract{
CTCV J2056--3014 is a nearby cataclysmic variable with an orbital
period of approximately $1.76$ hours at a distance of about $853$
light-years from the Earth. Its recently reported
X-ray properties suggest that J2056-3014 is an unusual
accretion-powered intermediate polar that harbors a fast-spinning
white dwarf (WD) with a spin period of $29.6$ s. The low X-ray
luminosity and the relatively modest accretion rate per unit area
suggest that the shock is not occurring near the WD surface. It has  been argued that, under
these conditions, the maximum temperature of
the shock cannot be directly used to determine the mass of the WD
(which, under the abovementioned assumptions, would be around $0.46$ $M_\odot$). Here, we
explore the stability of this rapidly rotating WD using a
modern equation of state (EoS) that accounts for electron--ion,
electron--electron, and ion--ion interactions. For this EoS, we determine the mass
density thresholds for the onset of pycnonuclear fusion reactions and study the impact of microscopic stability 
and rapid rotation on the structure and stability of WDs, considering them with helium, carbon, oxygen, and neon. From this analysis, we obtain a minimum mass for CTCV
J2056--3014 of $0.56~M_\odot$ and a maximum mass of around $1.38~M_\odot$.
If the mass of CTCV J2056--3014 is close to the lower mass limit, its
equatorial radius would be on the order of $10^4$~km due to rapid rotation. Such a radius is significantly larger than that of a nonrotating WD of average mass ($0.6\, M_\odot$), which is on the order of $7\times 10^3$~km. The effects on the minimum mass of J2056-3014 due to changes in the temperature and composition of the stellar matter were found to be negligibly small.}


   \maketitle
%
\section{Introduction} \label{introduction}

\cite{2020ApJ...898L..40L} report on X-ray Multi-Mirror
Mission (XMM-{\it Newton}) observations
that reveal CTCV J2056--3014 to be an X-ray-faint intermediate polar
harboring an extremely fast-spinning white dwarf (WD) with a coherent $29.6$ s
pulsation. This modulation seen in X-rays is also identified in all optical light curves. Over the last decade, there has been increasing interest
from the astrophysics community in fast-spinning WDs, both from the
theoretical and observational points of view. Typically, WDs rotate
with periods of days or even years.  A pulsating WD
named AR Scorpii was recently discovered
{\citep[see][]{2016Natur.537..374M,2016ApJ...831L..10G,2017NatAs...1E..29B}. It emits
electromagnetic radiation ranging from ultraviolet wavelengths into
the radio regime, pulsing in brightness with a spin period of $1.97$
min. Other sources have also been observed with similar
spin frequencies. A specific example is AE Aquarii, with a short rotation period
of $P=33.08$~s~\citep[see, e.g.,][]{2006A&A...445..305I,2008PASJ...60..387T,2014ApJ...782....3K}. The
pulsations were originally discovered in the optical
\citep{1979ApJ...234..978P}, then confirmed in soft X-rays
\citep{1980ApJ...240L.133P} and the ultraviolet
\citep{1994ApJ...433..313E}.  Furthermore, the XMM-{\it Newton} observatory has observed a WD
rotating faster than AE Aquarii.  It was shown
by~\citet{2009Sci...325.1222M} that the X-ray pulsator RX
J0648.0--4418, which belongs to the binary system HD
49798/RXJ0648.0--4418, is a massive ($M=1.28~ \textrm{M}_\odot$) WD with a very small spin period of $P=13.2$~s
\citep[see][]{1997ApJ...474L..53I}, but its nature is not clear (see \citealt{2016MNRAS.458.3523M} and \citealt{2018MNRAS.474.2750P} for details). More recently, \cite{2020arXiv200813242A} reported that,
based on Hubble Space Telescope ultraviolet data, V1460 Her exhibits strong pulsations with a period of $38.9$ s.

In addition, it is important to mention the newly discovered, highly magnetized, isolated, and rapidly rotating WD, designated ZTF J190132.9+145808.7, which, with a mass of $1.35M_\sun$ and a radius of $2140 ~\rm km$~\citep[see][]{2021Natur.595...39C}, is  as small as the Moon. Such a small radius implies that the star’s mass is close to the Chandrasekhar mass limit. Furthermore, since this WD is isolated, we are observing, for the first time, a very fast, nearby ($\sim 40$~pc) WD  that is not in a binary system but seems to have been originated by a merger of two WDs. In fact, this source may be the missing link to support a recent claim made in \citet{2012PASJ...64...56M} and \citet{2012IJMPS..18...96C,2014PASJ...66...14C} that some of the anomalous X-ray pulsars~\citep[AXPs; see][]{2014ApJS..212....6O,2017ARA&A..55..261K} that spin with period of $\sim $ 10 s could be isolated, very massive, magnetic, and fast WDs that resulted from WDs mergers \citep{2013ApJ...772L..24R}. Recently,  the quiescent spectral energy distribution of (AXP) 4U 0142+61 was reproduced with great success  from mid-infrared up to hard X-rays using plausible physical components and parameters~\citep[see][for details]{2017MNRAS.465.4434C,2020ApJ...895...26B}.

Several theoretical works regarding very magnetic, massive, and fast WDs have been published in the last few years ~\citep{2015NuPhA.937...17B,doi:10.1142/S021827181641025X,2016JCAP...05..007M,2020MNRAS.492.5949S,2020MNRAS.498.4426S}. More recently, \citet{10.1093/mnras/stz2734} showed that continuous gravitational waves can be emitted from rotating magnetized WDs and could possibly be detected in the near future by instruments such as Laser Interferometer Space Antenna (LISA), Big Bang Observer (BBO), and Deci-hertz Interferometer Gravitational wave Observatory (DECIGO), and the Einstein Telescope~\citep[see also][]{2020MNRAS.492.5949S,2020MNRAS.498.4426S}.

From a theoretical point of view, WDs can rotate at periods as short as $P\approx 0.5 \rm \; s$~\citep[see][]{2013ApJ...762..117B}, and they can be
formed, as we pointed out above, in double WD mergers~\citep[see][and references therein]{2012ApJ...749...25G,2013ApJ...772L..24R,2013MNRAS.428..579I,2018ApJ...857..134B}. \citet{livio_rotation_1998}
and \citet{livio_spins_1999} investigated the role of rotation for the maximum mass of a WD ~\citep[see also][]{1968ApJ...151.1075O,1968ApJ...151.1089O}. General relativistic effects on uniformly rotating WDs have been studied more recently by \cite{boshkayev_maximum_2013}.
Also, a general relativistic magnetohydrodynamic framework that describes rotating and magnetized axisymmetric WDs
was explored sequentially by \cite{10.1093/mnras/stv1983} and \cite{10.1093/mnras/stv2823}.

On the other hand, the two main observables of a WD, its mass and radius, both depend crucially on the equation of state (EoS). It
is worth mentioning that the mass presented by
\cite{2020ApJ...898L..40L} was poorly determined in the sense that it
is based on a fit that uses an X-ray spectrum model that may not be
the most appropriate for the source, in addition to other approaches
used to calculate the WD mass from the maximum temperature in the post-shock region. According to the authors, under such assumptions the mass would be around $0.46~M_\odot$. In addition, some arguments strongly suggest that the magnetic field of J2056 is very low~\citep[for details, see][]{2020ApJ...898L..40L}.

In this paper we describe WD matter in terms of helium, carbon, oxygen, or a mixture of oxygen ($64\%$) and neon ($36\%$), taking not only the electron Fermi gas contribution into account, but also the
contributions from electron--ion, electron--electron, and ion--ion
interactions. Recently, we performed a stability analysis of the matter
in the cores of WDs against pycnonuclear fusion
reactions~\citep[see][]{otoniel_strongly_2019,2021BrJPh..51..223M}. 
In the present paper,  we
determine theoretical bounds on the mass of CTCV J2056--3014 
that follow from mass shedding caused by rotation 
at the Kepler frequency 
and pycnonuclear fusion reactions among carbon nuclei in the
core of WD matter.

This paper is organized as follows. In Sect.~\ref{sec1} we present
the theoretical framework and methodology used to determined the WD
composition and structure, as well as details about Hartle's stellar rotation
formalism and pycnonuclear fusion reactions. In Sect.~\ref{sec2} we
present our main results. Concluding remarks are presented in
Sect.~\ref{sec3}.

\section{The model} \label{sec1}

The increase in more sensitive techniques of observation and detection
of WDs
\citep[see][]{eisenstein_catalog_2006,kepler/2013,2015MNRAS.446.4078K,2019A&ARv..27....7C},
coupled with advanced calculations of the properties of Fermionic
matter under extreme physical conditions, has led to considerable
interest in theoretical studies of the structure and evolution of
WDs~\citep[see, e.g.,][]{chamel_stability_2013,2013ApJ...762..117B,2014ApJ...794...86C,chamel_maximum_2014,2018ApJ...857..134B,otoniel_strongly_2019}. The
EoS of relativistic and degenerate WD matter, which accounts for the excess free energy of a one-component plasma
(OCP) made of ions derived from Monte Carlo simulations, is
computed from the Helmholtz free energy,
\begin{equation}
  F=F_{\mathrm{ id}}^{\mathrm{ ion}}+F_{\mathrm{ id}}^{(\rm e)}+F_{\rm e e}+F_{\rm i i}+F_{\rm i e} \, ,
\label{eq:F}
\end{equation}
where $F_{\mathrm{id}}^{\mathrm{ion}}$ denotes the free energy of a nonrelativistic classical gas given by
\begin{equation}
F_{\mathrm{id}}^{\mathrm{ion}}=N_{\rm j} k_{\rm B} T\left[\ln \left(n_{\rm j} \lambda_{\rm j}^{3} / g_{\rm j}\right)-1\right] .
\end{equation}
{Here, $N_{\rm j}=n_{\rm j} V$, $k_{\rm B}$ is the Boltzmann constant, $T$ the temperature of the gas, $n_{\rm j}$ the total number density of ions, and $\lambda_{\rm j}^{3}$  the thermal de Broglie wavelength, $\lambda_{\rm j}=\left(2 \pi \hbar^{2} / m_{\rm j} k_{\rm B} T\right)^{1 / 2}$, where $m_{\rm j}$ is the ion mass and $g_{\rm j}$ the spin multiplicity.
The free energy of electrons ($F_{\mathrm{id}}^{(\rm e)}$)
is given by~\citep[see][]{PhysRevE.58.4941}
\begin{equation}
F_{\mathrm{id}}^{(\rm e)}=\mu_{e} N_{e}-P_{\mathrm{id}}^{(\rm e)} V,
\end{equation}
where $\mu_{\rm e}$ is the electron chemical potential. 
The pressure ($P_{\text{id}}^{(\rm e)}$) and the electron number density ($n_{\rm e}=N_{\rm e} / V$) are functions of $\mu_{\rm e}$ and $T$ and can be written in terms of the Fermi-Dirac integrals $I_{\rm \nu}\left(\chi_{\rm e}, \tau\right),$ where $\chi_{\rm e}=\mu_{\rm e}/k_{\rm B} T$ and $\nu=1 / 2,3 / 2,$ and $5 / 2 .$ The chemical potential can be obtained  by
inverting the function $n_{\rm e}\left(\chi_{\rm e}, T\right)$ numerically.
The last three terms on the right-hand side of Eq.~(\ref{eq:F}) represent the free energies arising from electron--electron, ion--ion, and ion--electron interactions  given by ~\citep{PhysRevE.58.4941,potekhin_equation_2000}
\begin{eqnarray}
F_{\rm e e} &\equiv& f_{\rm e e} N_{\rm e} k_{\rm B} T, \\
F_{\rm i i} &\equiv& f_{\rm i i} N_{\mathrm{ion}} k_{\rm B} T,  \\
F_{\rm i e} &\equiv& f_{\rm i e} N_{\mathrm{ion}} k_{\rm B} T.
\end{eqnarray}The pressure ($P$) and the entropy ($S$) of a plasma with a fixed number of particles in a given volume ($V$) are obtained from the thermodynamic relations $P=-(\partial F / \partial V)_{T,N}$ and $S = -(\partial F  / \partial T)_{V,N}$, with  the internal energy of the system given by $U=F+T S$~\citep[see][]{PhysRevE.58.4941,potekhin_equation_2000}. 

We computed the EoS for a system of carbon ions at temperatures of $T=10^4$ K and $10^7$ K. Although there are no observational restrictions on what type of WD exists in the CTCV J2056--3014 system, it is worth mentioning that we did the calculations for different WD constitutions. The results are very similar if  heavier elements are considered, as shown in this paper. To be
specific, we described WD matter in terms of helium, carbon, oxygen, or a mixture of oxygen (64\%) and neon (36\%), taking the
contributions from electron--ion, electron--electron, and ion--ion interactions into account. Here, we used an EoS for the free energy of an ideal relativistic electron gas with an arbitrary degeneracy as well as for the free energy excess of the OCP. We used the Chabrier \& Potekhin EoS mainly because of the temperature dependence and electron interactions, which allowed us to create more realistic models to be compared with observed data. In Fig.~\ref{figEOS} we show the mass as a function of radius for WDs considering Chandrasekhar's EoS and the Chabrier \& Potekhin EoS for carbon and oxygen. It is important to mention that the Chabrier \& Potekhin EoS enables us to obtain different minimum masses according to the composition and temperature, in contrast to Chandrasekhar's EoS. Additionally, although the difference between the curves is quite small, the precision of the Chabrier \& Potekhin EoS is crucial since we are searching for minimum masses. An upper bound on the WD mass follows from the occurrence of electron capture and from the pycnonuclear fusion reaction instability (see Sect.~\ref{ssec:pycno}). These are marked with a solid square and a solid circle in Fig.~\ref{figEOS}. We thus conclude that the upper limit on the gravitational mass of J2056--3014 is around $1.38 \, M_\odot$, provided it is a carbon or oxygen WD rotating near its mass-shedding limit.

\begin{figure}[h]
\includegraphics[width=8cm]{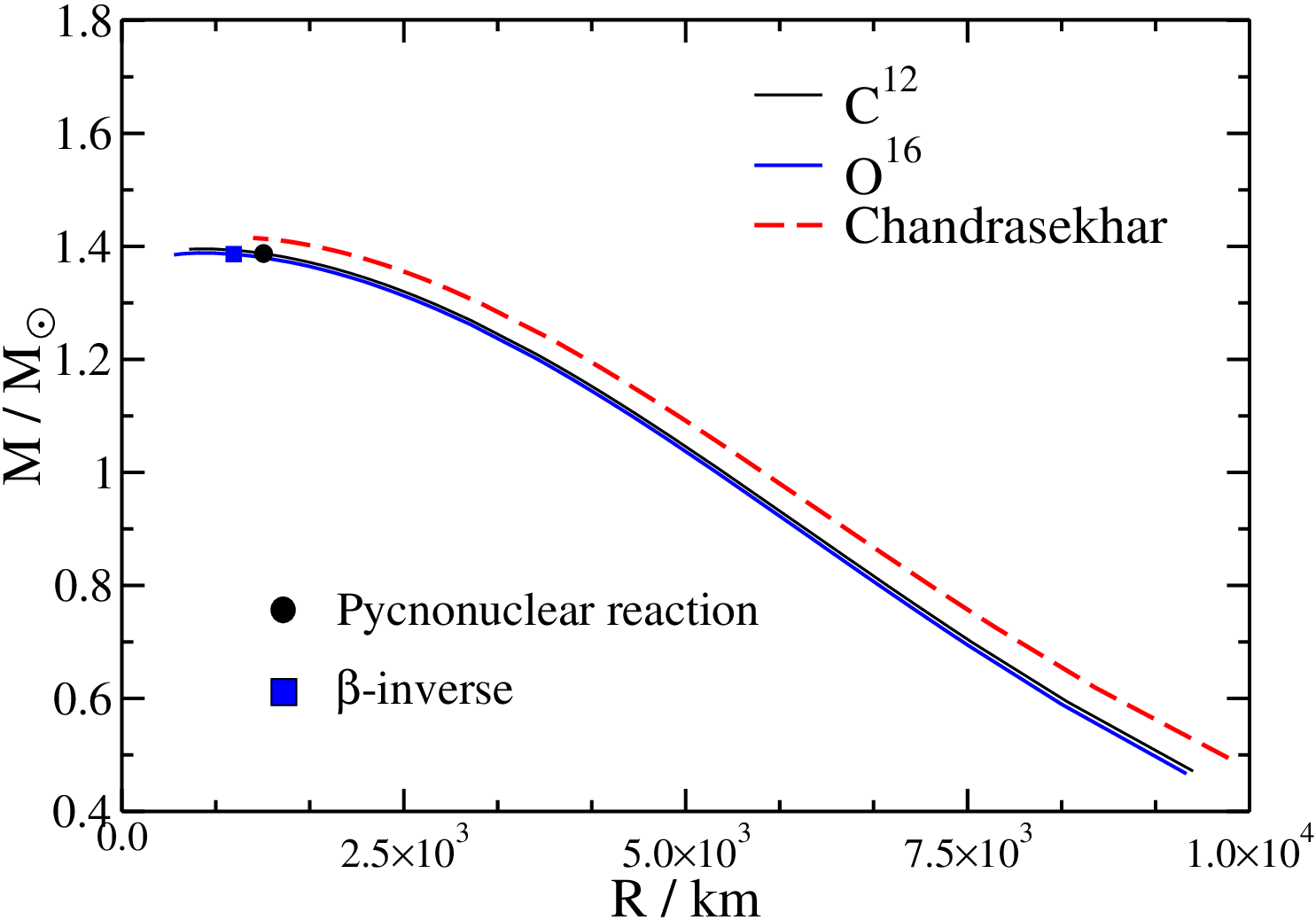}
\caption{Mass–radius relationships of WDs computed 
for the EoSs of Chabrier \& Potekhin (black and blue curves) and Chandrasekhar (dashed red curve).}
\label{figEOS}
\end{figure}
 
\subsection{Hartle's stellar rotation formalism }\label{ssec:hartle}

The structure of WDs is governed by hydrostatic equilibrium, where
gravity is balanced by the outward pressure generated by a
relativistic electron gas. Recently, it has been noted that
contributions from general relativity ought to be taken into
account when modeling the structure of
WDs~\citep[see][]{boshkayev_maximum_2013}. The
Tolman-Oppenheimer-Volkoff equation describes the structure of
nonrotating stellar objects and, in particular, determines the radii
and masses of such objects. However, in order to account for
rotational effects, Einstein's field equations need to be solved for a
metric that accounts for rotational deformation and the dragging of
local inertial
frames~\citep[see][]{1967ApJ...150.1005H,1968ApJ...153..807H,Friedman1986}. Such
a metric is given by
\begin{equation}
  ds^2 = - e^{2 \nu} dt^2 + e^{2 \psi} (d\phi-\omega dt)^2 + e^{2\mu}
  d\theta^2 + e^{2\lambda} dr^2 \, ,
\end{equation}
where the metric functions $\nu$, $\psi$, $\mu$, and $\lambda$ and the
frame dragging frequency ($\omega$) depend on the radial coordinate ($r$), the polar angle ($\theta$), and, implicitly,  the star's rotational
frequency ($\Omega$; see \citealt{Friedman1986,weber_pulsars_1999} for details). Thus, we performed a 2D calculation where the rotational star deformation is described by the polar angle ($\theta$), and the compact objects are symmetric around the axis of rotation. The $\Omega$ is taken to be
in the range of $0 \leq \Omega \leq \Omega_K$, where $\Omega_K$
$(=2\pi/ P_{K})$ denotes the Kepler (mass-shedding) frequency, which
terminates stable rotation. The  $\Omega_K$ sets an absolute upper limit
on rapid rotation. The Kepler frequency must be computed 
self-consistently, together with Einstein's field equations for the
metric functions, from
\begin{equation}
\Omega_{\rm K} = \omega + \frac{\omega'} {2\psi'} + e^{\nu-\psi}
\sqrt{\frac{\nu'}{\psi'} + \left(\frac{\omega'}{2\psi'} e^{\psi-\nu}
  \right)^2} \, ,
\label{Kepler}
\end{equation}
where the primes denote partial derivatives of the metric functions
with respect to the radial
coordinate~\citep[][]{1992ApJ...390..541W,PhysRevD.50.3836}. Hartle's perturbative treatment of compact objects leads to results that are in good agreement with those obtained by a numerically exact treatment of Einstein's field equations. This is particularly the case for the mass increase due to rapid rotation at the Kepler frequency~\cite[][]{1992ApJ...390..541W}.

\begin{table}
\caption{Theoretically established mass ($M$), radius ($R$), and central mass density ($\rho_{\rm c}$)  of J2056--3014 set by mass shedding (rotation at the Kepler period of 29.6~s) and pycnonuclear fusion reactions among carbon nuclei in the core of a WD. These
apply to CTCV J2056--3014 if this compact object is a pure carbon WD. }            
\label{table1}      
\centering                          
\begin{tabular}{c c c c}        
\hline\hline                 
Process & $M/M_\odot$ & $R/\rm km$ & $\rho_c/\rm g\; cm^{-3}$ \\    
\hline                        
   Rotation at Kepler period  & $0.56$ & $10965$ & $1.70\times10^6$\\
   Pycnonuclear reactions &$1.38$ & $1308$ & $9.26\times10^9$ \\ 
\hline                                   
\end{tabular}
\end{table}

\subsection{Pycnonuclear fusion reactions}\label{ssec:pycno}

We recently performed a stability analysis of the matter in the
cores of WDs against pycnonuclear fusion reactions and electron capture reactions~\citep[see][for details]{otoniel_strongly_2019,2021BrJPh..51..223M}. In
the current paper we investigate the stability of carbon WD matter to pycnonuclear fusion
reactions using up-to-date theoretical
models~\citep[see][]{gasques_nuclear_2005,golf_impact_2009}. We assumed a uniform nuclear composition 
throughout the star and focused on nuclear fusion reactions only\footnote{It is known that, at
  sufficiently high densities in the interior of WDs, the inverse
  $\beta-$decay, or electron capture, process becomes energetically
  favorable. Such a process destabilizes the star because the electrons
  provide all the pressure required to balance gravity.  However, as
  pointed out by~\citet{otoniel_strongly_2019}, these densities are
  higher than the density at which pycnonuclear fusion reactions in a
  $^{12}$C WD set in.}  among heavy atomic nuclei, schematically
expressed as $^{A}_{Z} X + ^{A}_{Z} X \rightarrow
^{2A}_{2Z} Y$.  Examples of such a reaction are carbon on carbon ($^{12}\textrm{C}+^{12}\textrm{C}$) and oxygen on oxygen ($^{16}\textrm{O}+^{16}\textrm{O}$). 
Pycnonuclear reactions have been theoretically calculated over a significant range of stellar densities~\citep[see][]{gasques_nuclear_2005}, including the density ranges that exist in the interiors of WDs \citep[see][]{chamel_stability_2013,chamel_maximum_2014,10.1093/mnras/stx781}. The nuclear fusion rates at which pycnonuclear reactions may actually proceed are highly uncertain because of some poorly constrained parameters~\citep[see][and references therein]{gasques_nuclear_2005,yakovlev_fusion_2006}.

\begin{table*}[h!]
 \caption{Minimum mass ($M$), equatorial radius ($R$), and central mass density ($\rho_c$) predicted for CTCV J2056--3014 for different matter compositions and temperatures if the star's rotational period of $29.6$~s is the Kepler period.} \label{tab:2}
\centering
\begin{tabular}{c c c c}
\hline
 Constitution and Temperature & $M/M_\odot$ & $R\rm/km$  & $\rho_c/\rm g\;cm^{-3}$\\ 
 \hline  
 $^{4}_2\rm{He}$, $T=10^7 \rm \; K$ & 0.65 & 11196 &$2.70\times10^6$\\
    $^{12}_6\rm{C}$, $T=10^4  \rm \; K$ & 0.56 & 10965 & $1.70\times10^6$ \\ $^{12}_6\rm{C}$, $T=10^7 \rm \; K$&0.57 & 11181 &$2.08\times10^6$ \\
    $^{16}_8\rm{O}$, $T=10^4 \rm \; K$&0.55 & 10964 &$1.64\times10^6$ \\
  $^{16}_8\rm{O}$, $T=10^7 \rm\; K$&0.55 & 11036 &$1.68\times10^6$ \\
   $^{12}_6\rm{O}(64\%)+^{20}_{10}\rm{Ne}(36\%),~T=10^4 \rm \;K$&0.53 & 10774 &$1.44\times10^6$ \\
  $^{12}_6\rm{O}(64\%)+^{20}_{10}\rm{Ne}(36\%),~T=10^7 \rm\; K$&0.54 & 11188 &$1.79\times10^6$ \\
  \hline  
\end{tabular}
\end{table*}

\section{Stability boundaries} \label{sec2}

In this section we establish constraints on the mass and radius of the
rapidly rotating WD CTCV J2056--3014 that follow from stable rapid
stellar rotation and the microscopic stability of matter, as described
in Sects. \ref{ssec:hartle} and \ref{ssec:pycno}.
Figure~\ref{figPM} shows the rotational periods of carbon WDs spinning
at their respective mass-shedding periods ($P_K$).  The solid square in
this figure corresponds to a WD whose Kepler period is equal to the
rotation period observed for CTCV J2056--3014 of $P= 29.6$~s. If
this period is close to the mass-shedding period, the lower bound on
the gravitational mass of J2056--3014 would be $0.56\, M_\odot$
and the central mass density of this object would be $1.70\times10^6$~${\rm g\; cm^{-3}}$ (see Table~\ref{table1}). An upper bound on the mass follows
from the occurrence of pycnonuclear reactions in WDs. This is the case
for the WD model marked with a solid triangle on the dashed line. We
thus conclude that the upper limit on the gravitational mass of J2056--3014 is around $1.5\, M_\odot$, provided it is a carbon WD
rotating near its mass-shedding limit. The green shaded area in
Fig.~\ref{figPM} shows the location of WDs spinning below that limit.

When heavier elements, such as oxygen and a mixture of oxygen ($64\%$) and neon ($36\%$), are considered, our calculations indicate that the minimum WD mass for CTCV J2056--3014 decreases only slightly. 
Temperature effects were also considered on the minimum mass star, and we obtain a small increase in the equatorial radius with an increase in temperature. In our calculations, the difference between the equatorial and polar radius of rapidly spinning
WDs was found to be as large as around 10\% (for the star with $0.56$ M$_{\odot}$ and $P=29.6$ s). In addition, the temperature effect in the minimum mass is also negligible (see~Table~\ref{tab:2}).

\begin{figure}[h]
\includegraphics[width=8cm]{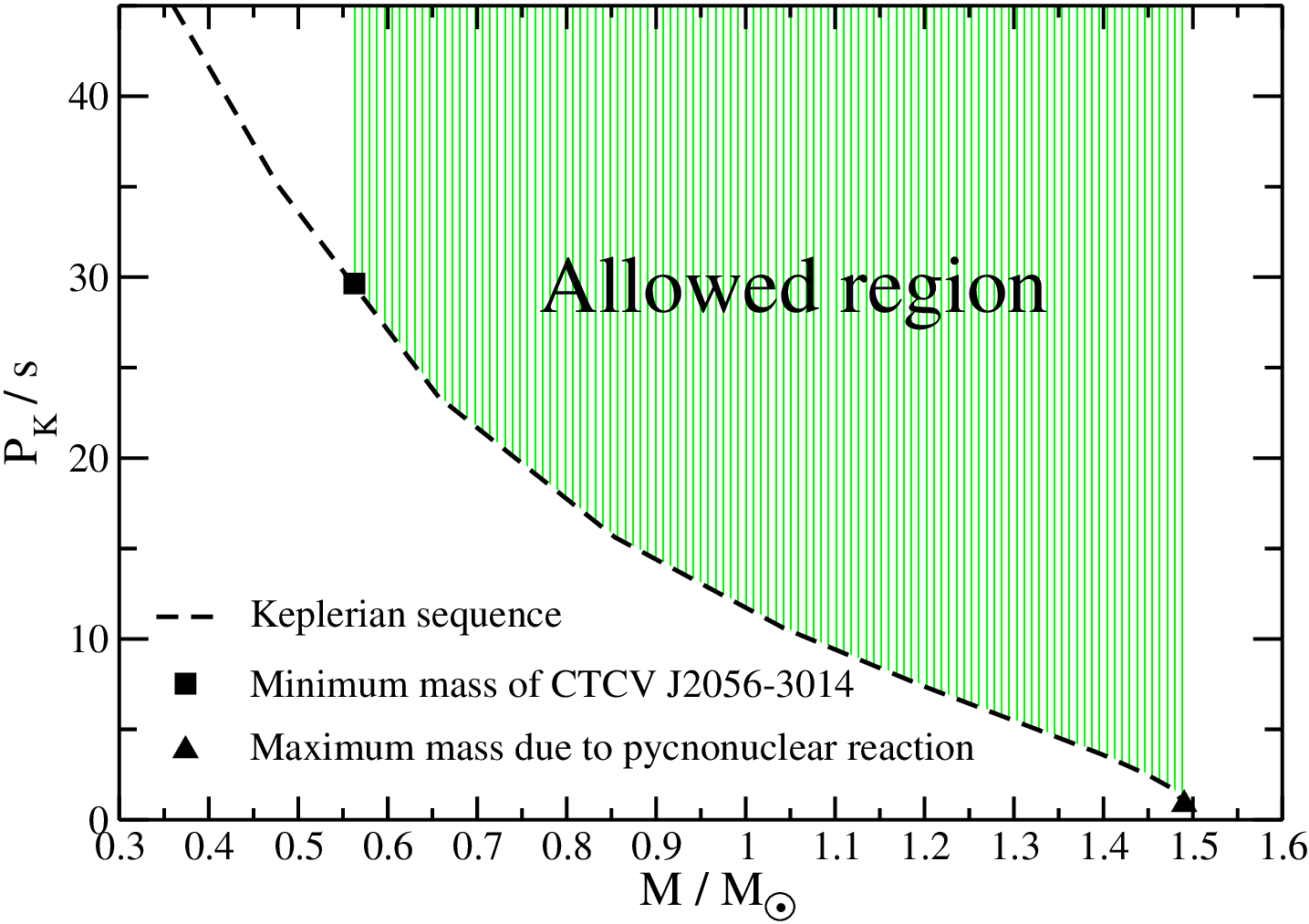}
\caption{Mass limits 
on CTCV J2056--3014 determined by rotation at the Kepler  period ($P_K$) and
the onset of pycnonuclear reactions (for carbon WD). A minimum mass (solid square) of $0.56\, M_\odot$ is obtained if CTCV J2056--3014's rotational period of $29.6$~s is close to the mass-shedding limit. The upper
mass limit (solid triangle) of around $1.5\, M_\odot$ results from the onset of pycnonuclear reactions if CTCV is a carbon WD.}
\label{figPM}
\end{figure}

In Fig.~\ref{figMrho} we show the gravitational WD mass as
a function of central mass density.  In contrast to
Fig.~\ref{figPM}, where all stars along the dashed line rotate at
their respective mass-shedding limits, the stars along the dashed line
shown in this figure all rotate at  $29.6~\rm s$, the spin period of CTCV
 J2056--3014. This lies well above the respective
Kepler periods for most stars along the sequence. The most massive WD
on the sequence, marked with a solid triangle, has a central density
of $9.26\times 10^9~{\rm g \;cm^{-3}}$ (see~Table~\ref{table1}).  At this
density, pycnonuclear reactions set in and terminate the microscopic
stability of WDs.  The area highlighted in green shows the mass-central
density region predicted to exist in a WD with a rotation period of $P\geq 29.6 \; \rm s$. We stress that this is an important theoretical 
limit that will be useful for
the interpretation of WDs discovered
in the future.

The reaction rates used to study the pycnonuclear reactions were
calculated with an astrophysical S factor computed for the NL2 nuclear
model parametrization~\citep[for details, see][]{otoniel_strongly_2019}. The threshold density at which
pycnonuclear reactions are triggered in a carbon WD is shown in
Table~\ref{table1}. We also adopted the pycnonuclear reaction timescale of $10 \rm\; Gyr$ to obtain the threshold density. It is worth mentioning that the
reaction rates are rather uncertain and that the analytic astrophysical
S factor has an uncertainly of  about $3.5$, which affects
the density thresholds of pycnonuclear reactions and their reaction
times considerably.

\begin{figure}[h]
\includegraphics[width=8cm]{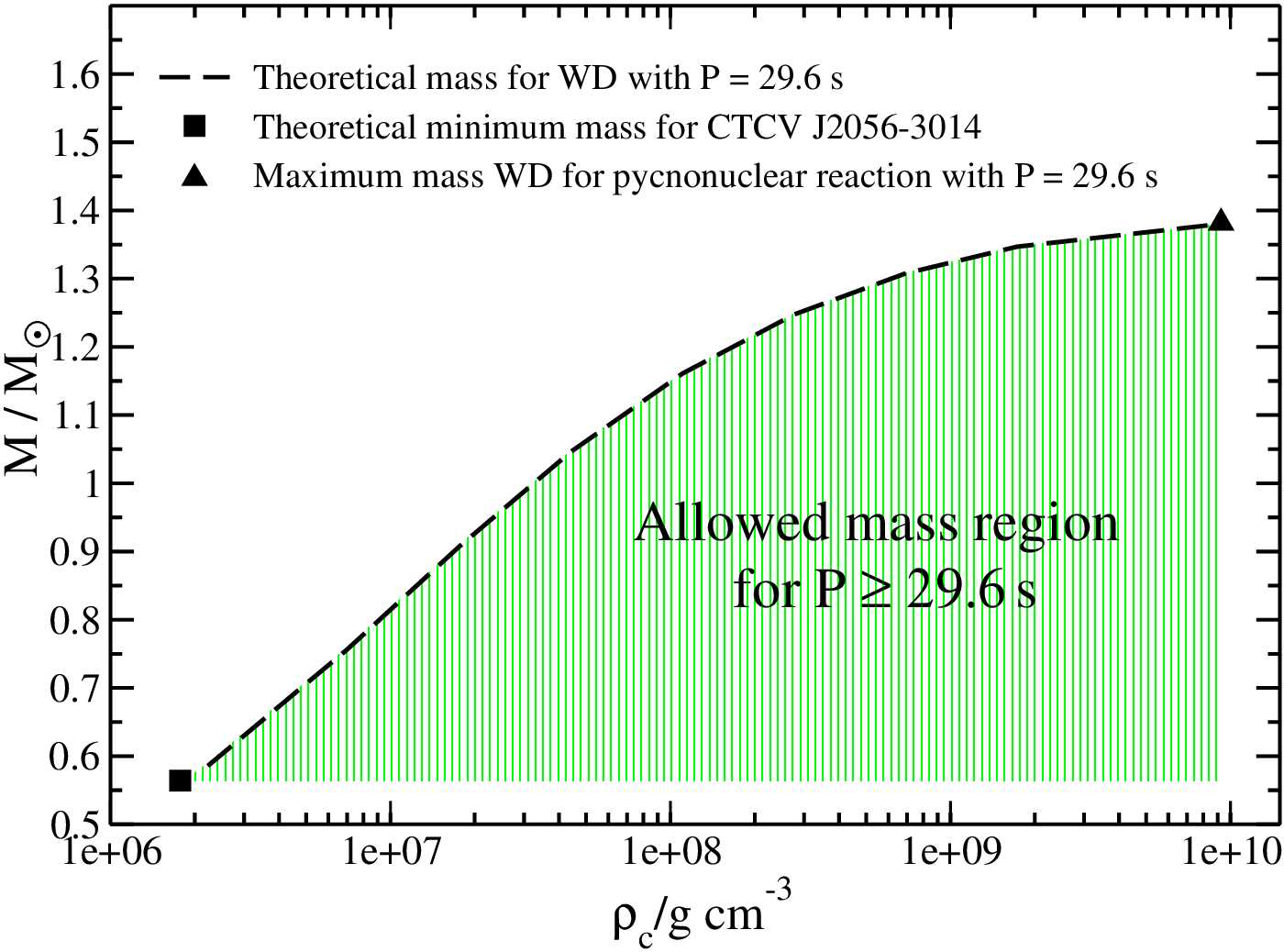}
\caption{Carbon WD mass vs. central mass density. All stars along the dashed curve rotate at $P=29.6$~s, as observed for J2056--3014. The solid square marks the least massive WD, whose Kepler period is 29.6~s. For all other stars along the dashed line, the rotation period of $29.6$~s is greater than the  
Kepler period. The most massive WD on the sequence, marked with a solid triangle, has a central density of $9.26\times10^{9}~{\rm g\; cm^{-3}}$. At this density, pycnonuclear reactions set in and microscopic stability ends.
\label{figMrho}}
\end{figure}

It is important to mention what happens with the allowed mass region in Fig.~\ref{figMrho} if different constitutions of matter are considered. As reported by \citet{1961ApJ...134..669S}, the threshold density for pycnonuclear reactions is smaller for helium than for carbon. Indeed, the threshold density for helium is around $87\%$ smaller than for  carbon. This would make  the  allowed mass regions for He WDs smaller than those of the carbon WDs shown in Fig.~\ref{figMrho}. Besides, as can be observed in Table~\ref{tab:2}, the allowed region of helium WDs  decreases even more at the minimum mass. Therefore, for oxygen and even a mixture of oxygen ($64\%$) and neon ($36\%$), the stability region would be greater than for a pure carbon WD.

\section{Concluding remarks}\label{sec3}

In this paper we stress that some macro and micro physical aspects, such as rotation and pycnonuclear fusion reactions, are of great relevance for the self-consistent description of the structure and the assessment of
the stability of CTCV J2056--3014. As argued by~\citet{2020ApJ...898L..40L}, the mass inferred for this object was poorly estimated in the sense that it is based on the assumptions
often made when solving the structure equations in the X-ray-emitting region that is produced by a shock near the WD. These assumptions
include radial accretion, free-fall from infinity, and a shock near the WD surface such that the maximum temperature of the shock cannot be directly used to calculate the mass of the WD.  Here, we used the EoS of  helium, carbon, oxygen, and neon WDs that accounts for electron--ion, electron--electron, and ion--ion
interactions. For this EoS, we determined the mass
density thresholds for the onset of pycnonuclear fusion reactions and studied the impact of microscopic stability 
and of rapid rotation on the structure and stability of WDs. Our analysis predicts a minimum mass for CTCV
J2056--3014 of $0.53~M_\odot$ for a  mixture of oxygen ($64\%$) and neon ($36\%$).
We estimate the theoretical limit of the mass-central density region for future observations of
 WDs with rotation periods $P\geq 29.6$s.  Should the mass of CTCV J2056--3014 be close to the lower mass limit at $T=10^4$ K, its equatorial radius would be on the order of $10^4$~km  due to rapid rotation. Such a radius is significantly greater than the radius of a nonrotating WD of average mass ($0.6\, M_\odot$), which is on the order of $7\times 10^3$~km.

Also, we stress an important point of our analysis. Very fast WDs with periods in the range of tens of seconds, and even a WD pulsar, have been detected in 
recent  years. In order to spin so fast, such
WDs need to be  massive, with the minimum mass limit determined by the Kepler frequency.
The Zwicky Transient Facility is currently discovering large numbers of very massive and fast-spinning WDs~\citep[see][]{2021Natur.595...39C}. These and future observations will help us to better understand such objects.

\begin{acknowledgements}
The authors thank the referee for comments which helped to improve the  quality of the manuscript. We are deeply grateful to Claudia V. Rodrigues for useful discussions.
E.0.\ is grateful for the support of Pr\'{o}-Reitoria de Pesquisa e
Inova\c{c}\~{a}o - EDITAL 01/2020/PRPI/UFCA. J.G.C. is likewise grateful to the support of  CNPq (421265/2018-3 and 305369/2018-0), FAPESP Project No. 2015/15897-1, and NAPI "Fenômenos Extremos do Universo" of Fundação de Apoio à Ciência, Tecnologia e Inovação do Paraná. M.M.\ acknowledges financial support from FAPESP under
the thematic project 13/26258-4, Capes, CNPq and INCT-FNA (Proc. No. 464898/2014-5). S.P.N. thanks Conselho Nacional de Desenvolvimento Cient\'ifico e Tecnol\'ogico (CNPq), Grant No.       $140863/2017-6$ and CAPES for the financial support. F.W.\ is supported by the National
Science Foundation (USA) under Grant PHY-1714068 and PHY-2012152.
     
\end{acknowledgements}

%
\bibliographystyle{aa} 
\bibliography{ref} 
%

\end{document}